\documentclass[aps,prl,twocolumn,showpacs,groupedaddress]{revtex4}
\usepackage{graphicx}

\begin{document}


\title{A Molecular Micromaser}
\author{Chris P. Search, Weiping Zhang, and Pierre Meystre}
\affiliation{Optical Sciences Center, The
University of Arizona, Tucson, AZ 85721}

\date{\today}

\begin{abstract}
We show that photoassociation of fermionic atoms into bosonic molecules 
inside an optical lattice can be described using a Jaynes-Cummings Hamiltonian 
with a nonlinear detuning. Using this equivalence to the Jaynes-Cummings dynamics, 
we show how one can construct a micromaser for the molecular field in each lattice
site.
\end{abstract}

\maketitle

The recent experiment by Greiner {\em et al.} that produced a
Mott-insulator transition in ultracold $^{87}$Rb confined
in an optical lattice \cite{greiner} has stimulated a great deal
of interest in strongly correlated systems of ultracold atoms in
optical lattices \cite{jaksch2,oosten}. At the same time, there
has been a growing interest in the possibility of forming a
molecular Bose-Einstein condensate (BEC) via either photoassociation
\cite{julienne,heinzen} or a Feshbach resonance. To date,
experiments have demonstrated the formation of large numbers of
molecules via photoassociation of an $^{87}$Rb BEC \cite{wynar}
and using a Feshbach resonance applied to an $^{85}$Rb BEC
\cite{wieman} and a degenerate Fermi gas of $^{40}$K \cite{jin}.

These two lines of research have recently converged in theoretical proposals to 
photoassociate bosonic atoms in the Mott-insulator state into
molecules \cite{jaksch1,damski,moore,molmer,esslinger}. Photoassociation in optical lattices 
offers distinct advantages over quasi-homogenous systems. In 
particular, one can obtain occupation numbers per lattice site that are of order unity,
thereby minimizing the losses due to inelastic collisions. Secondly, the 
center-of-mass states of the atoms and molecules are restricted to the lowest energy
Wannier state at each lattice site, thereby avoiding the difficulties associated 
with a multimode problem \cite{goral}.

In this letter we extend this work to the case of fermionic atoms. Our central result is that as 
a consequence of Pauli statistics,  this problem reduces mathematically within each lattice site 
to a situation almost identical to that of the familiar micromaser 
\cite{micromaser_th,micromaser_ex} of quantum optics, the only significant difference 
resulting from the presence of an intrinsic nonlinearity in the case of matter-wave fields. 
It is well established that the radiation field generated in a micromaser exhibits 
a number of quantum mechanical features that are absent in normal 
lasers and masers. For example, micromasers exhibit strong sub-Poissonian 
statistics \cite{rempe} and even the ability to dynamically generate Fock states \cite{scully}. 
Unlike lasers, which approach a classical coherent state far above threshold, 
micromasers undergo multiple phase transitions characterized by sharp changes in the photon 
statistics from sub-Poissonian to super-Poissonian \cite{guzman}. These unique features  
find their origin in the ``granular'' nature of the radiation
field, combined with the lack of noise, such as spontaneous emission, which washes 
out the effects of the quantum-mechanical atom-field dynamics in conventional lasers.

The close analogy between the situation at hand and the micromaser therefore 
opens up exciting novel avenues to generate non-classical states of molecular
fields and to study new types of {\em dynamical} quantum phase transitions 
in matter wave systems. More generally, it permits the extension of the numerous, well-established
applications of cavity QED, to matter-wave optics. Of particular interest in this
context is the fact that optical lattices automatically lead to the realization of an {\em array}
of individually addressable molecular micromasers. 

We proceed by giving a broad overview of the theory of 
molecular micromasers. We  discuss under which conditions
the problem of fermionic photoassociation in a lattice can be mapped to the 
micromaser problem, and then present some of the important physical 
implications of this result.

We consider a 2-D optical lattice in the $xy$-plane that confines 
fermionic atoms of mass $m_f$ in two hyperfine states $|1\rangle$ and $|2\rangle$. 
The atoms are coupled to a molecular boson of mass $m_b=2m_f$ via
a two-photon stimulated Raman transition
\cite{julienne,heinzen,wynar}. The optical lattice potential $V_f$
seen by the atoms is $V_f({\bf
r})=\sum_{\xi=x,y}V^{(f)}_{\xi}\cos^2k_{\xi} 
\xi+\frac{1}{2}\kappa_z^{(f)}z^2$ where $V^{(f)}_{\xi}>0$ and we have 
included a transverse confining potential along the $z$-axis. A similar potential, 
with $f \rightarrow m$, confines the molecules. We assume that the system 
is at zero temperature and that the filling factor --- the number of
fermions of each type that occupy every lattice site --- is
$n_F\leq 1$ at all times. Hence the fermions only
occupy the lowest Bloch band of the lattice. For deep lattices, for
which the energy separation between the first and second Bloch
bands is much greater than the atom-molecule interaction energy,
the molecular state formed by photoassociation has likewise a center-of-mass
wave function in the first Bloch band \cite{jaksch1}.

The Hamiltonian for the atom-molecule system in the first Bloch
band of the lattice is $\hat{H}=\sum_i 
\left( \hat{H}_{0i}+\hat{H}_{Ii} \right)+\hat{H}_T$
where,
\begin{eqnarray}
&&\hat{H}_{0i} = 
\hbar(\omega_b+\delta)\hat{n}_{bi}+\hbar\omega_f (\hat{n}_{1i}+\hat{n}_{2i}) \nonumber \\
&&+ \hbar \frac{1}{2}U_{b} 
\hat{n}_{bi}(\hat{n}_{bi}-1)+\hbar U_{x}\hat{n}_{bi}(\hat{n}_{1i}+\hat{n}_{2i})+\hbar U_{f}\hat{n}_{1i}\hat{n}_{2i}, \nonumber \\
&&\hat{H}_{Ii} =\hbar\chi(t) b^{\dagger}_ic_{1,i}c_{2,i}+H.c.,
\end{eqnarray}
and $H_T$ accounts for tunnelling between lattice sites. 

Here, $b_i$ is the bosonic annihilation operator for molecules in the Wannier
state of the lowest Bloch band centered at the lattice site $i$, $w_b({\bf r}-{\bf r}_i)$,
and similarly, $c_{\sigma,i}$ is the fermionic annihilation operator for state $|\sigma\rangle$ with the
Wannier wave function $w_f({\bf r}-{\bf r}_i)$. The corresponding number operators 
$\hat{n}_{bi}=b^{\dagger}_ib_i$ and $\hat{n}_{\sigma i}=c^{\dagger}_{\sigma,i}c_{\sigma,i}$ have
eigenvalues $n_{bi}$ and $n_{\sigma i}$, respectively.
The terms proportional to $U_{b}$, $U_{x}$, and $U_{f}$ describe the 
on-site two-body interactions between molecules, atoms and
molecules, and atoms, respectively. 

The interaction Hamiltonian, $\hat{H}_{Ii}$, describes the conversion of atoms into ground-state molecules
via two-photon stimulated Raman photoassociation. $\chi(t)$ is proportional to the far 
off-resonant two-photon Rabi frequency associated with two nearly co-propagating lasers with frequencies
$\omega_{1}$ and $\omega_{2}$ \cite{wynar}. Note also that the matter fields have been written in an interaction 
representation in which the molecular field is rotating at the frequency difference of the two
lasers, so that $\delta=\nu_m-(\nu_1+\nu_2)-(\omega_{1}-\omega_{2})$ where
$\hbar\nu_{\sigma}$ and $\hbar \nu_m$ are the internal energies of
the atoms and molecules, respectively.

We assume in the following that the lattice is sufficiently deep that atomic 
tunnelling can be ignored during the time intervals $\tau$ when the photoassociation 
lasers are on. We further assume that the molecules are in the
Mott-Insulator state, for which tunnelling is completely suppressed for the duration of 
the experiment \cite{oosten,albus}. This second assumption is essential in the realization 
of a molecular micromaser, but we emphasize that in the open system that we consider in the 
following, it does not imply that the molecules  at a given site are constrained to be in a 
Fock state. Consistently with these assumptions, we ignore $\hat{H}_T$ altogether in the following. 

In the absence of intersite coupling, we need only consider the Hamiltonian
$\hat{h}=\hat{H}_{0i}+\hat{H}_{Ii}$ at a single lattice site.
In what follows, we drop the lattice site label for notational clarity.
We then proceed by introducing the mapping \cite{anderson},
\begin{eqnarray}
\sigma_{-}&=&c_1c_2, \quad \sigma_{+}=\sigma_{-}^{\dagger}=c_2^{\dagger}c_1^{\dagger} \nonumber \\
\sigma_{z}&=&c^{\dagger}_1c_1+c^{\dagger}_2c_2 -1
\label{mapping}
\end{eqnarray}
where $\sigma_{+}=|e\rangle\langle g|$ and
$\sigma_{-}=|g\rangle\langle e|$ are the raising and lowering
operators for a fictitious two-state system, and $\sigma_z=|e\rangle\langle
e|-|g\rangle\langle g|$ is the population difference between its
upper and lower states. Here, $|e\rangle
=c^{\dagger}_2c^{\dagger}_1|0\rangle$ and $|g\rangle=|0\rangle$.
We note that this mapping only holds if $c_1$ and $c_2$ are 
fermionic operators, hence, our subsequent discussion would not hold for bosonic atoms. 
The mapping (\ref{mapping}) allows us to reexpress $\hat{h}$ exactly as,
\begin{eqnarray}
\hat{h}&=&\hbar\left(\omega_b+U_{x}\right)\hat{n}_b+\hbar\left(\omega_f+
U_{x}\hat{n}_b\right)\sigma_z \nonumber \\
&+&\hbar\left(\chi(t)b^{\dagger}\sigma_
{-}+ \chi^*(t)b\sigma_{+} \right) +\frac{\hbar}{2}U_{b}\hat{n}_b(\hat{n}_b-1) 
\label{JCH}
\end{eqnarray}
where we have dropped constant terms and made the redefinitions 
$\omega_b+\delta\rightarrow \omega_b$ and $\omega_f+U_{f}/2\rightarrow 
\omega_f$. 

In the limit $U_{b},U_{x}\rightarrow 0$, this Hamiltonian reduces to the Jaynes-Cummings 
model of interaction between a quantized, single-mode electromagnetic field
and a two-level atom. This model is a cornerstone of quantum optics \cite{scully}. 
Because it is exactly solvable, it permits the understanding of detailed 
aspects of the dynamics of light-matter interaction. The development of ultra-high Q cavities and 
cavity QED has led to the experimental realization of systems that closely approach the
ideal situation of the Jaynes-Cummings model. Of particular relevance in the present context
is the micromaser, which consists of an ultra high-$Q$ microwave cavity in which two-level atoms are
injected one at a time. 

The dynamics of the micromaser are governed by three mechanisms: the injection of a sequence
of individual atoms from a very dilute atomic beam inside the microwave cavity, the Jaynes-Cummings
interaction between these atoms and the cavity mode, and cavity dissipation, which 
can normally be neglected while an atom transits through the cavity, but that dominates the field dynamics
in the intervals between atoms. A similar situation can be achieved in the present 
system: all that is required is to inject a sequence of pairs of fermionic atoms inside the lattice well, 
turn on the photoassociation lasers for some time interval $\tau$ to 
introduce a Jaynes-Cummings-like dynamics --- 
where the electromagnetic field mode is replaced by the molecular field --- and finally turn off these fields 
for a time $T$ and let dissipation take over. This sequence is then repeated 
to build up the molecular field. The close analogy with the micromaser
makes it clear that this ``molecular micromaser'' 
can generate strongly nonclassical molecular fields, and, more
generally, opens the way to "cavity atom optics.''    

We now discuss the three key mechanisms that govern the ``molecular micromaser'' 
dynamics, starting with the Jaynes-Cummings-like evolution.
For $\chi=const$, equation (\ref{JCH}) can be solved just like the Jaynes-Cummings Hamiltonian 
within the two-state manifold $\{ |e,n_b\rangle , |g,n_b+1\rangle \}$. The dynamics is then in the
form of quantized Rabi oscillations between the two states at the frequency
\begin{equation}
{\mathcal 
R}_{n_b}=\sqrt{\left[2\omega_f-\omega_b+\left(2U_{x}-U_{b}\right)n_b\right]^2+4|\chi|^2(n_
b+1)}.
\end{equation}
In particular, if the system is initially prepared in the state 
$|e,n_b\rangle$, the probabilities for the system to be in the two states of the 
manifold after a time $\tau$ are $\left| c_{e,n_b}(\tau)\right| 
^2=1-C_{n_b+1}(\tau)$ and $\left| c_{g,n_b+1}(\tau)\right| ^2=C_{n_b+1}(\tau)$ where
\[
C_{n}(\tau)=\frac{4|\chi|^2n}{ {\mathcal R}_{n-1}^2}\sin^2\left(\frac{1}{2} {\mathcal 
R}_{n-1} \tau \right).
\]
Note that unless $U_{b}=2U_{x}$, the detuning in ${\mathcal R}_n$ depends on the number of 
molecules present and thus resonant Rabi 
oscillations are only possible for a single manifold. 

The pump mechanism consists of continuously injecting
pairs of fermionic atoms into each lattice site during the interval $T$ when $\chi=0$.
This can be realized in a 2-D lattice by a beam of fermions 
incident on the lattice along the $z$-axis in internal states $|f_{\sigma}\rangle$ 
that have polarizabilities opposite to those of $|\sigma=1,2\rangle$. This 
results in them seeing a repulsive instead of an attractive potential centered at $z=0$. 
The energy of the incident beam is taken to be much less than the barrier height, so 
that tunnelling is negligible. 
Two lasers directed along the lattice can then be used to stimulate Raman 
transitions from the untrapped to the trapped states, $|f_{\sigma}\rangle \rightarrow 
|\sigma\rangle$. The coupling constants are then given by 
$i\kappa_k=\int d^3{\bf r}e^{ikz}\Omega_{2}({\bf r})w_{f}^*({\bf 
r})/2\sqrt{V}$ where $\Omega_{2}$ is the two-photon Rabi frequency and $V$ 
the normalization volume for the plane wave states of the incident beam \cite{jaksch2}.

For a broadband continuum of reservoir states, one can derive a Born-Markov master 
equation for the density matrix of the trapped fermions, 
\begin{eqnarray}
\left. \frac{\partial\rho(t)}{\partial t} \right|_{\rm pump} &=&-\frac{\Gamma}{2}\sum_{\sigma=1,2}\left(
\bar{n}\left[ c_\sigma c^{\dagger}_\sigma \rho-2c^{\dagger}_\sigma \rho c_\sigma+\rho
c_\sigma c_\sigma^{\dagger} \right]  \right . \nonumber \\
&+& \left . (1-\bar{n})\left[ c^{\dagger}_\sigma c_\sigma\rho -2c_\sigma\rho c^{\dagger}_\sigma+\rho
c^{\dagger}_\sigma c_\sigma \right] \right) \label{pump}
\end{eqnarray}
where $\bar{n}$ is the number of incident fermions with energy $\hbar k^2/2m_f=\omega_f$ and
the ``pumping rate'' $\Gamma=2\pi|\kappa(\omega_f)|^2 D(\omega_f)$ with $D(\omega)$ 
being the density of states for the plane waves \cite{input-output}. In order to pump the
lattice site with fermions we take $\bar{n}=1$. Since the two fermionic
species evolve independently in Eq. (\ref{pump}), we can express
their density matrix as the tensor product of the
density operator for each species, $\rho^{(1)}(t)\otimes
\rho^{(2)}(t)$. Eq. (\ref{pump}) then has the solution
$\rho^{(\sigma)}_{0,0}(t)=\exp(-\Gamma t)\rho^{(\sigma)}_{0,0}(0)$ and
$\rho^{(\sigma)}_{1,1}(t)=1-\rho^{(\sigma)}_{0,0}(0)\exp
(-\Gamma t)$ where $\rho^{(\sigma)}_{0,0}=\langle
0|\rho^{(\sigma)}|0\rangle$ and $\rho^{(\sigma)}_{1,1}=\langle
0|c_\sigma\rho^{(\sigma)}c^{\dagger}_\sigma|0\rangle$.
For $\Gamma t\gg 1$, $\rho^{(\sigma)}_{1,1}(t)=1$, i. e. the
fermions are in the state $|e\rangle=c^{\dagger}_2c^{\dagger}_1|0\rangle$ with unit
probability.

Finally, the damping of the molecular mode is described via
a phenomenological master equation that includes the dominant loss mechanisms.  
They include in particular three-body inelastic collisions between a
molecule and two atoms in which the atoms form a dimer with the
resultant loss of the atoms and molecule. This occurs
at a rate $\gamma_3 \sim \hbar 
a_{f}^4/m_f(\ell^{(f)}_{x}\ell^{(f)}_{y}\ell^{(f)}_{z})^2$ where $a_{f}$ is 
the atomic scattering length and $\ell^{(f)}_{\xi}$ the characteristic spatial extent of 
$w_f({\bf r})$ along the three axes \cite{jack,fedichev}. Another loss mechanism is 
spontaneous emission from the molecular excited state from which the optical lattice 
and photoassociation beams are detuned. When $\chi(t)=0$, these losses are due solely 
to scattering of lattice photons. This gives a rate 
$\gamma_1\approx\gamma_e\omega^{(b)}/4|\Delta|$ for 
a blue detuned lattice, using a harmonic approximation for the optical potential 
at the lattice nodes. Here $\gamma_e$ is the excited state linewidth and 
$\Delta$ the detuning from the excited state. The contribution
to the master equation due to both loss mechanisms is
\begin{equation}
\left. \frac{\partial\rho(t)}{\partial t} \right|_{\rm loss}
=-\frac{\gamma_1}{2}\left[b^{\dagger}b\rho-b\rho b^{\dagger}\right]
- \frac{\gamma_3}{2}\left[B^{\dagger}B\rho-B \rho B^{\dagger}\right] +H.c. \label{decay1}
\end{equation}
where $B=bc_1c_2=b\sigma_{-}$ \cite{jack}. For molecules created in
their rotational-vibrational ground state, we can ignore losses due to rotationally or 
vibrationally inelastic collisions between molecules, or between atoms 
and molecules \cite{heinzen}. 

In the limit of strong pumping, $\Gamma \gg \gamma_{1,3}$, the
atoms reach their steady state before
the state of the molecules has noticeably changed. We can then
replace Eq. (\ref{decay1}) with a coarse-grained master equation 
valid for time intervals much larger than $\Gamma^{-1}$. To do this we
substitute the steady-state values of Eq.
(\ref{pump}) into the density operator,
$\rho(t)=\rho^{(b)}(t)\otimes |e\rangle \langle e|$ where
$\rho^{(b)}$ is the molecular density operator, and then
trace over the states of the atoms in Eq. (\ref{decay1}). This yields finally
\[
\frac{\partial\rho^{(b)}(t)}{\partial t} =-\frac{1}{2}\gamma\left[b^{\dagger}b\rho^{(b)}-2b\rho^{(b)}
b^{\dagger}+\rho^{(b)} b^{\dagger}b \right] \equiv \mathcal{L}[\rho^{(b)}] 
\]
where $\gamma=\gamma_1+\gamma_3$. We note that for $a_f\sim 100a_0$ and $m_f\sim 10a.m.u.$ one has 
$\gamma_3^{-1}\sim 10s$ while $\gamma_1^{-1}\sim 1s$ for a detuning of $10000$ line 
widths and $\omega^{(b)}\sim 2\pi\times 10^{4}s^{-1}$.

Just like in micromaser theory, the weak damping of the molecular mode allows us to 
assume that during the intervals $\tau$ of photoassociation 
the evolution is unitary and given by $\hat{h}$. Since $\tau\sim 1/\sqrt{n+1}|\chi|$ for
complete Rabi oscillations to occur in the presence of $n$ molecules, this implies that the
photoassociation fields must satisfy $\sqrt{n+1}|\chi|\gg \gamma_{1,3}$. The resulting molecular
density operator is then
\begin{equation}
\rho^{(b)}(t_l+\tau)={\rm Tr}_{\rm atoms}\left[e^{-i\hat{h}\tau}\rho(t_l)e^{i\hat{h}\tau}\right]\equiv
F(\tau)[\rho(t_l)]
\end{equation}
where $t_l=l(T+\tau)$ for $l=0,1,2,...$. In order to determine $F(\tau)$ explicitly, we 
recall that at the end of each interval $T$ every lattice site is occupied by a pair 
of fermions with unit probability for $\Gamma T\gg 1$. Since the fermions
are uncorrelated with the state of the molecules already present,
the reduced density matrix at $t_l+\tau$ is then
$\rho^{(b)}(t_l+\tau)=\sum_n p_n(t_l+\tau) |n\rangle\langle n|$ where
$p_n(t_l+\tau)=[1-C_{n+1}(\tau)]p_n(t_l)+C_n(\tau)p_{n-1}(t_l)$. Here
$p_n=\langle n|\rho^{(b)}|n \rangle$ and we have assumed that
$\rho^{(b)}$ is initially diagonal in the number state basis.

The formal solution for the molecular density operator is then
$\rho^{(b)}(t_{l+1})=\exp({\mathcal L}T)F(\tau)\rho^{(b)}(t_l)$.
We note that $T$ is controlled by the timing of the photoassociation 
beams, in contrast to conventional micromasers where $T$ corresponds to the arrival
times of the atoms and is therefore normally a random variable. 

The steady-state molecular field is given by
the return map condition $\bar{\rho}^{(b)}=\rho^{(b)}(t_{l+1})=\rho^{(b)}(t_l)$.
If the interval $T$ is such that $\Gamma^{-1}\ll T \ll \gamma^{-1}$,
we can expand the exponential in the formal solution to lowest order in
${\mathcal L}$. The steady-state condition is then
${\mathcal L}[\bar{\rho}^{(b)}]=T^{-1}(1-F(\tau))\bar{\rho}^{(b)}$,
which has the same form as the steady state solution of the
micromaser if one identifies $T^{-1}$ with the rate
at which atoms are injected into the cavity \cite{micromaser_th}. The steady state solution 
in the number-state basis is,
\begin{equation}
\bar{p}_n=\bar{p}_0\prod_{l=1}^{n}\frac{C_l(\tau)}{\gamma Tl} 
\label{statistics}
\end{equation}
where $\bar{p}_n$ is the probability of having $n$ molecules and
$\bar{p}_0$ is determined by the conservation of probability, $\sum_n
{\bar p}_{n}=1$. Eq. (\ref{statistics}) has the same form as the zero-temperature 
micromaser photon statistics, except that
$C_l(\tau)$ contains an $l$-dependent detuning in ${\mathcal R}_l$ 
resulting from the two-body interactions.

Figures 1 through 3 highlight important features of the molecule statistics as a function
of $N_{ex}= 1/\gamma T$, the number of pumping cycles per lifetime of 
the molecules, and of the dimensionless strength for the 
nonlinear detuning $\beta=(2U_{x}-U_{b})/2|\chi|$. The micromaser pump 
parameter is given by $\Theta=(N_{ex})^{1/2}\chi \tau$. Figure 1 shows the 
average number of molecules at each site, $\langle \hat{n}_b \rangle$, as a 
function of $\Theta$ for the linear resonance condition, 
$\eta\equiv(2\omega_f-\omega_b)/2|\chi|=0$.
We see that the lasing threshold is not affected by 
the nonlinear detuning and occurs at $\Theta\approx 1$. However, above 
threshold $\langle \hat{n}_b \rangle$ is strongly suppressed as $\beta$ is 
increased. This is because the effective detuning between the fermion 
atom pairs and molecules increases with increasing molecule number due to 
the self-phase modulation of the molecules, $U_{b}\hat{n}_b(\hat{n}_b-1)$, and 
the AC Stark shift of the atoms, $U_{x}\hat{n}_b\sigma_z$. For $\beta>0(<0)$ this can be 
compensated for by choosing a finite negative (positive) detuning, 
$\eta<0(>0)$, as shown in Fig. 2.

\begin{figure}
\includegraphics*[width=8cm,height=4cm]{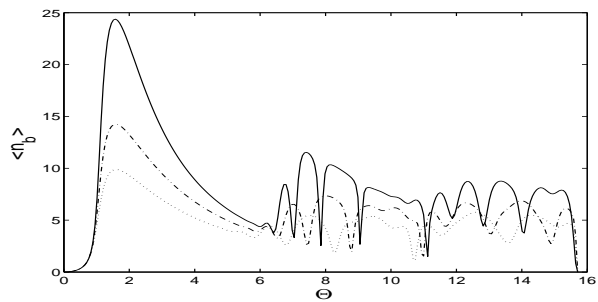}
\caption{Average number of molecules, $\langle n_b \rangle$, versus $\Theta$ for 
$N_{ex}=25$, $\eta=0$, and 
$\beta^2=0$(solid line), $0.05$ (dashed dot), $0.15$ (dotted line). }
\label{fig1}
\end{figure}

\begin{figure}
\includegraphics*[width=8cm,height=4cm]{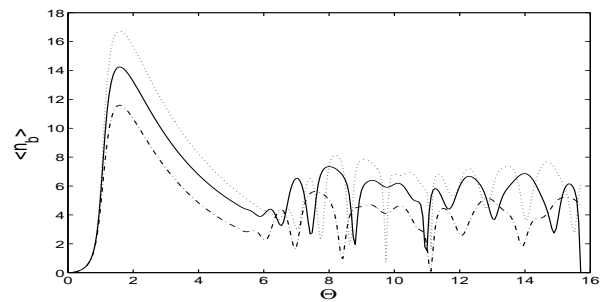}
\caption{$\langle n_b \rangle$ versus $\Theta$ for $N_{ex}=25$ and 
$\beta=\sqrt{0.05}$ for $\eta=0$
(solid line), $+1$ (dashed dot), $-1$ (dotted line). }
\label{fig2}
\end{figure}

Figure 3 shows the Mandel $Q$ parameter, $Q=\left(\langle \hat{n}_b^2\rangle -\langle 
\hat{n}_b\rangle^2\right)/\langle \hat{n}_b\rangle -1$. Just above threshold, 
the molecule distribution is strongly super-Poissonian ($Q>0$) and then become 
sub-Poissonian ($Q<0$) until $\Theta\approx 2\pi$ after which $Q$ shows very 
sharp oscillations. We note that the number fluctuations decrease with 
increasing $|\beta|$ and show smaller super-Poissonian peaks. The sharp 
resonances in $Q$ and $\langle \hat{n}_b \rangle$ for large $\Theta$ are 
attributable to downward trapping states \cite{scully}. Unlike the 
conventional micromaser, there is no thermal noise present in the damping of 
the molecular field. If we replace $\mathcal{L}[\rho^{(b)}]$  with the 
master equation for coupling to a finite temperature bath with a thermal 
occupation of $\bar{n}_b$, then the rapid fluctuations in 
$Q$ are averaged out. 

\begin{figure}
\includegraphics*[width=8cm,height=4cm]{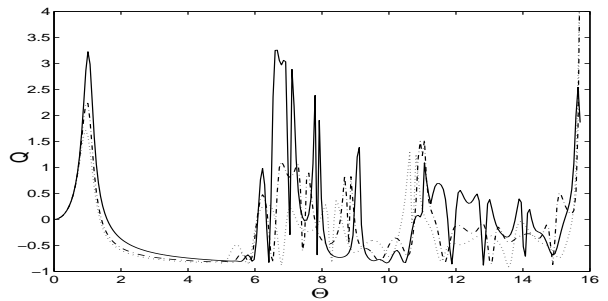}
\caption{$Q$ versus $\Theta$ for $N_{ex}=25$, $\eta=0$, and 
$\beta^2=0$(solid line), $0.05$ (dashed dot), $0.15$ (dotted line). }
\label{fig3}
\end{figure} 

In conclusion, we have examined the photoassociation of fermionic atoms into 
bosonic molecules and shown that the dynamics can be described by an effective 
Jaynes-Cummings Hamiltonian with a nonlinear detuning resulting from the two 
body elastic interactions of the molecules. We have also presented the theory 
of a molecular micromaser.

This work is supported in part by the US Office of Naval Research,
by the NSF, by the US Army Research Office, by NASA, and
by the Joint Services Optics Program. C. P. S. would like to thank B. P. Anderson
for helpful discussions.

\end{document}